# Hydrated Silicate Ionic Liquids: Ionic Liquids for Silicate Material Synthesis

*Dries Vandenabeele[a], Nikolaus Doppelhammer[a,b], Sambhu Radhakrishnan[a,c], Vinod Chandran C[a,c], Wauter Wangermez[a], Anjul Rais[a], Elisabeth Vandeurzen[a], Bernhard Jakoby[b], Christine Kirschhock[a], Eric Breynaert[a,c,*]*

a. Centrum voor Oppervlakte chemie & Katalyse: Karakterisatie en Applicatie Team (COK-KAT), KU Leuven.
b. Institute for Microelectronics and Microsensors - Johannes Kepler University Linz, Altenberger Straße 69, 4040 Linz.
c. NMRCoRe - NMR/X-Ray platform for Convergence Research, KU Leuven, Celestijnenlaan 200F Box 2461, 3001-Heverlee, Belgium.
* Corresponding Author: Eric.Breynaert@kuleuven.be

**ABSTRACT**

The development of Hydrated Silicate Ionic Liquids, which are hypo-hydrated room temperature melts of alkali silicates, has created new opportunities for synthesizing porous silicates. Their discovery also allows to reinterpret the role of an ionic liquid for zeolite synthesis and offer unique opportunities to study crystallization in situ. The reduced complexity of HSILs renders them excellent model systems yielding a large space of zeolite phases and framework compositions, and even allowing computer simulations. This work summarizes the impact of alkalinity, hydration, and cation type on silicate speciation and provides physicochemical data spanning relevant temperatures and compositions for zeolite formation. The data is meant as a reference point to further accommodate and verify (computational) models.

**IONIC LIQUIDS FOR INORGANIC POROUS MATERIALS**

Ionic liquids (ILs) most commonly occur as organic liquids, popular as green alternatives for volatile and/or harmful organic solvents. They are liquid salts and are thus exclusively composed of ions. Even though any molten salt in principle fits this definition, the term 'ionic liquid' usually refers to salts that are liquid below 100°C.[1] Even though the score of possible cation-anion combinations exceeds $10^8$, most literature in the field of ionothermal synthesis of (porous) materials focuses on organic imidazolium-derived cations, regardless of the type of targeted porous material.[2–6]

For the synthesis of zeolites and zeotype materials, ionic liquids act simultaneously as solvents and as structure-directing agents.[2] In the absence of water, there is no competition between water molecules and cations at the growing surface, enabling 'true' structure-directing.[2] The role of the anionic part of the ionic liquid is less clear. Mostly, weakly coordinating anions are explored, such as halides, hydroxides, or complex (organic) molecules.[7,8] By definition, weakly coordinating anions do not associate strongly with the cation, allowing the cation to maximally coordinate with framework elements and thus maximally act as a template. The anion plays a critical role in shaping the properties of the ionic liquid, influencing factors such as melting point and hygroscopicity. While these changes in chemical properties can significantly impact the final structure—such as whether the anion is occluded within the framework—its effect is considered inductive rather than templating or structure-directing.[2]

Ionothermal synthesis with organic ILs has been very successful for aluminophosphates, delivering a range of known and new topologies.[2,5,6,9,10] For aluminosilicate synthesis the approach has been problematic, very often requiring the addition of water, hydroxides, or fluorides to overcome the low solubility of silicate.[8] In such cases, the role of IL is mostly reduced to that of a regular structure-directing agent. Some reports also have demonstrated fluoride-catalyzed IL degradation during the synthesis, further complicating the interpretation of their role in the synthesis.[5,9,11] Aside from the traditional organic ILs, also silicate-based fully inorganic ionic ILs exist. They represent a radical innovation of the use and role of ionic liquids in zeolite formation: Rather than functioning as a structure-directing agent, the hydrated silicate ionic liquids (HSILs) directly serve as the source of framework elements, much like is often the case for coordination polymers.[2,12] HSILs can be considered as liquid hydrous alkali hydroxy-silicate salts.[12] The high alkalinity and low water content force the silicate into small oligomers, associated with hypo-hydrated alkali cations. HSILs are not the first molten salts to incorporate water; the earliest examples can be traced back to Barrer's work in the 1940s.[13–15] The classification of HSILs as IL is valid because the water is fully occupied in the coordination sphere of the ions, ensuring it is unavailable to act as a solvent.[12,13,15] While HSILs are liquid at room temperature, alkali silicate counterparts such as sodium metasilicate hydrates are solid at ambient temperature and only melt at elevated temperatures (72 and 47°C for respectively $Na_2SiO_3 \cdot x H_2O$ with x = 5 and 9).[16]

Pursuing to improve the understanding of HSILs on a molecular level, Vekeman and coworkers developed a molecular model using MD simulations.[17] The modeled conductivity shows a remarkable qualitative agreement with experimental conductivity values enabling the assessment of important parameters such as ion association and its structuring role. Further development of such models relies on the availability of data describing micro- and macroscopic properties. This manuscript aims to provide such macroscopic data, focusing on conductivity, density, and viscosity covering relevant compositions and temperatures. Additionally, we summarize previously communicated results and address remaining gaps to provide an overview of the HSILs' microscopic properties.

**MATERIALS AND METHODS**

1. *HSIL synthesis*

Following the synthesis protocol developed by Francis Taulelle and coworkers, HSILs are easily obtained by controlled hydrolysis of tetraethyl orthosilicate (TEOS) under agitation in an alkali hydroxide solution.[12,13] During the hydrolysis, a liquid-liquid phase separation occurs, yielding a water-ethanol mixture on top of a more dense, mostly ionic HSIL liquid. Virtually all the silicates and cations are converted into hypo-hydrated alkali cation silicate ion pairs, residing in the bottom phase. Ethanol and bulk water are mainly found in the lower-density phase on top. Synthesis protocols have been developed for Na, K, and Cs, with starting compositions designed to provide a broad range of homogeneous compositions. Recipes and final compositions are given in Table 1. HSILs remain stable for years and do not crystallize with increased pressure and/or temperature. As-made HSILs can be mixed with additional components, such as additional cations and water, to obtain the range of compositions suitable for zeolite synthesis. Note that upon high dilution with water, these systems should no longer be considered Hydrated Ionic Liquids. Previous work has defined the transition from HSIL to solution as the abrupt appearance of colloidal species, observed around the conductivity maximum, occurring around 15 $H_2O$/MOH for all tested cations and alkalinites.[18,19]

Table 1: HSIL recipes and final compositions.

|    | Alkali Source | TEOS* mol (g) | MOH mol (g) | $H_2O$ mol (g) | Final HSIL composition |
|----|---------------|---------------|-------------|----------------|------------------------|
| Na | >97%, ACROS Organic | 1 (47.23g) | 1 (8.95g) | 24.5 (100g) | $NaSiO(OH)_3 \cdot 3.2\ H_2O$ |
| K  | >85%, Sigma-Aldrich | 1 (47.23g) | 1 (14.69g) | 20.5 (80g) | $KSiO(OH)_3 \cdot 4.6\ H_2O$ |
| Cs | 99.5% $CsOH.H2O$, Sigma-Aldrich | 1 (47.23g) | 1 (37.07g) | 8.7 (31.49g) | $CsSiO(OH)_3 \cdot 3.2\ H_2O$ |

* 98 %, Acros organics

2. *Characterization*

$^{29}$Si NMR spectra of the HSIL samples were acquired in a Bruker 300 MHz Avance III spectrometer equipped with a 10mm Si-free BBO probe. The HSIL samples were filled in a 5mm sapphire tube (Wilmad-LabGlass), part of a high-pressure NMR setup allowing in situ zeolite synthesis.[20] The spectra were acquired with 3072 transients, which was further referenced to primary reference with a relaxation delay of 20 s, and 90° radio-frequency pulse of 25 kHz at 25 °C. $^1$H decoupling with Waltz64 sequence with an RF pulse of 5 kHz was applied during acquisition. The spectra were referenced to 0.1 M solution of Sodium trimethylsilylpropanesulfonate (DSS) in $D_2O$, which was further referenced to primary reference, tetramethylsilane (TMS).

Density was recorded using an Anton Paar DSA 5000 M. Viscosity measurements were conducted on a Haake Mars 3 Rheometer (Thermo Fisher Scientific) with C60/1° Ti L cone-plate cylinder geometry. Viscosity was measured at 10 logarithmically spaced shear rates in the upward direction within the range of 5 to 500 $s^{-1}$.

Conductivity was measured using Differential Impedance Spectroscopy (DIS) in a custom-built setup.[21–24] In contrast to conventional impedance spectroscopy, this method eliminates many spurious effects thus enabling highly accurate conductivity measurement in highly conducting, harsh media.[21–24] The AC voltage was adjusted for the conductivity of the sample between 0.015 and 0.1V. Spectra were recorded in downward direction using two settings: 5 equally spaced interelectrode distances between 4 and 6 cm and 31 logarithmically spaced frequencies between 1 kHz and 1 MHz per electrode distance (method 1) and 11 equally spaced electrode distances between 4 and 7 cm, using 20 logarithmically spaced frequencies between 10 kHz and 1 MHz (method 2). The respective methods are indicated in the table provided in the Harvard database. Each reported conductivity value represents the average of at least three individual measurements. The (experimental) conductivity data reported by Vekemans et al are also included in the analysis. The reader is referred to the original paper for the experimental details.[17]

Synchrotron diffraction patterns were recorded at the BM26 at the ESRF in Grenoble.[25] Sample loading and pressurization are described in Vandenabeele et al.[26] The sample was kept at temperature using a cryostream and rotated between 0 and 180° during data acquisition to ensure homogenous heating. A beam energy of 20 keV was used to overcome the high absorption of Cs and to achieve a reasonable signal-to-noise ratio. The data was collected on a Pilatus 300K-WF detector mounted at 28.2 cm to the center of the capillary, covering a Q-range of 1.2-8.45 $Å^{-1}$. Azimuthal integration was performed with ESRF's in-house Bubble software.[27] The patterns were recorded for 15 seconds each. The recorded intensities were normalized for the varying beam intensity.

**MICROSCOPIC PROPERTIES OF HYDRATED SILICATE IONIC LIQUIDS**

The molecular structure of HSILs is the result of a complex set of interactions between water, silica, and alkali cations. In as-made HSILs, the liquid structure is water scarce so the structure and properties are heavily influenced by ion association.[28–30] Although ion association is a typical feature for ionic liquids, it deserves special attention, as it crucially affects the crystallization of porous silicates.[18,29,31–33] Pellens and coworkers hypothesize that ion association is essential for crystallization directly from a liquid phase.[18] This section also highlights the impact of compositional parameters on silicate speciation, usually studied via $^{29}$Si NMR. While individual species sometimes stand out, most are part of a complex and dynamic mixture, making it close to impossible to identify each species individually. Speciation is therefore often simplified to "connectivity," $Q^n$. This parameter indicates the average number of T-atoms (Al or Si, expressed as n) connected to a silicon center through siloxane bonds. Figure 1 shows typical ranges for each connectivity type, as assigned by Hendricks, as well as some exemplary species.[34]

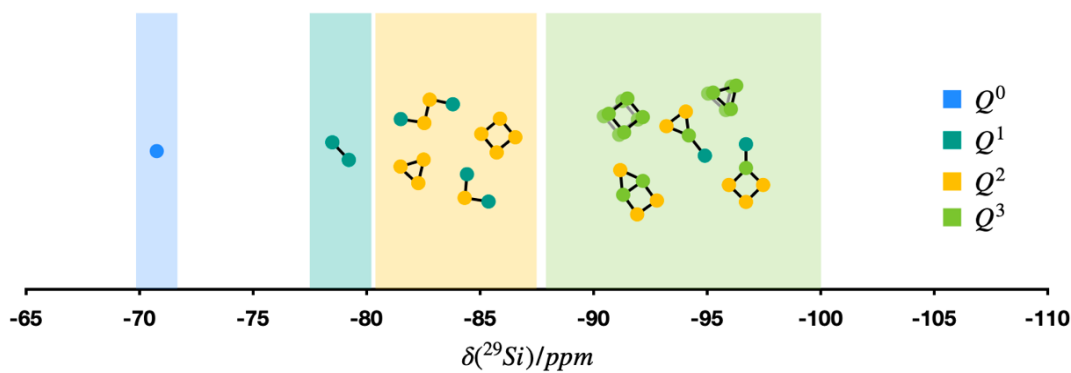

**Figure 1: Typical ranges for each connectivity type. The figure also provides some examples of species and the connectivity of their silicon atoms.**

Before exploring these properties in detail, we confirm that the HSIL is not in a metastable state by submitting an HSIL with composition $Cs_2SiO_2(OH)_2 \cdot 11\ H_2O$ to 296 K at 5 bar for 10 min and to 500 K at 170 bars of $N_2$ pressure for 45 min while assessing the diffraction pattern. As can be seen in Figure 2, the increase in pressure and temperature does not trigger crystallization. The diffraction intensity increases without significantly changing the profile. The absence of crystallization indicates that HSILs are stable, even in conditions that are severe for our purposes.

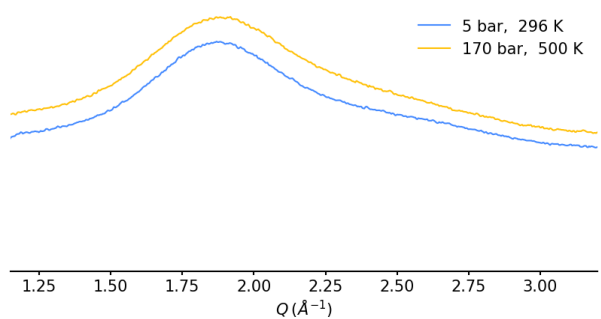

**Figure 2: The diffraction pattern of $Cs_2SiO_2(OH)_2 \cdot 11\ H_2O$ was recorded under two different conditions. The absence of Bragg reflections indicates that the HSIL is stable, even in severe conditions for our purposes.**

*Cations*

While literature describing liquid speciation of dilute (gel-like) silicates is prevalent[28,29,31–36], data for HSILs are scarce.[18,37–39] In general, alkali cations are mainly characterized by their size which affects their charge density and polarizability. In the abundance of water, this leads to promoting (kosmotropic, e.g. Li$^+$ or Na$^+$) or disrupting (chaotropic, e.g. K$^+$ and Cs$^+$) effects on the H-bonding network of water. Kosmotropes are hard ions, having a high charge density and a tightly bound inner hydration shell. In systems where water is deprived, ion hydration competes with ion association, with silicates in this instance, to stabilize the ions.[17,29] Chaotropes are characterized by a more labile hydration shell. Even though their interaction is weaker because of their larger ionic radius, it has been hypothesized that their higher polarizability allows them to stabilize larger silicate ions in silicate solutions.[29] Nevertheless, the overall impact of the type of alkali cation on the silicate connectivity is minimal.[34,40] Both Na and Cs have on average about 2.2 neighboring T-atoms on each silicon center for a solution with 1 mol% Si and Si/OH = 1.[34] McCormick and coworkers also found that the pH of the solution is roughly independent of the cation type so that also the average silicate ionization is minimally impacted by the cation type.[36]

In material sciences, organic cations like tetra-alkyl amines (TAA) are known as structure-directing agents. TAAs affect silicate speciation more significantly than alkali cations.[34] For example, tetra-methyl amide (TMA) strongly promotes the formation of double four-rings, while tetra-ethyl amide (TEA) also stabilizes double three-ring and six-ring structures.[34,41–44] All OSDAs are classified as chaotropic.[45]

The nature of the inorganic cation does not significantly affect the average connectivity in HSIL systems either. Figure 2 provides the $^{29}$Si NMR spectra for HSILs with the composition MSiO(OH)$_3 \cdot$ 6 H$_2$O, where M represents Na, K, or Cs. The fit parameters and the resulting difference spectra are given in Figure SI 1 and Table SI 1. The average connectivity was found to be around 2.3 for all investigated cations. We find an overall slight decrease in the Q$^3$ contribution relative to Q$^2$ species. The monomeric fraction, located between -70 and -72 ppm, is low, approximately 1% for the three tested HSIL systems. Noteworthy is that the peak position gradually moves towards more negative chemical shifts with increasing cation size. Smaller cations have a stronger electron-withdrawing effect, which deshields the siloxanes more effectively than larger cations, thus shifts the peak position towards more negative values.[29,36]

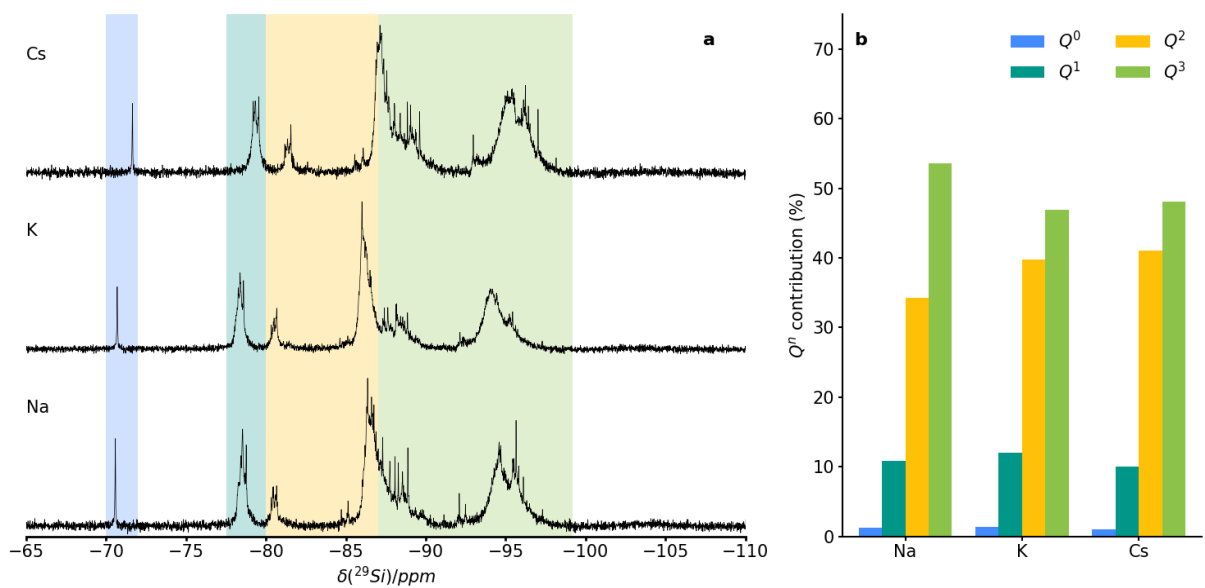

**Figure 2 (a) $^{29}$Si spectra for HSILs with composition MSiO(OH)$_3$.6 H$_2$O. The peaks are categorized by their connectivity type, leading to the average connectivity values presented in (b).**

*Alkalinity*

Alkalinity, commonly expressed as Si/OH, crucially affects silicate speciation.[36] On the molecular level, silicate chemistry is governed by the deprotonation of silicates, forming water and consuming hydroxides, and oligomerization (oxolation). Exemplary reactions are given in Figure 3. For the hyperalkaline, hypo-solvated alkali silicates under investigation, the Si/OH ratio is limited to 1 as gel formation is triggered above that stoichiometric ratio.[13]

Si/OH ratios close to unity are characterized by complex speciation, ranging from simple oligomers to larger double-ring structures. When the alkalinity increases, the average nuclearity decreases, leading to progressively smaller oligomers until only monomeric silica remains.[35–37,46] Silicate deprotonation is additionally promoted in water-deprived media as it generates water. The higher average charge per anion repulses other negatively charged silicates while the amount of available leaving groups decreases; OH is a better leaving group in the oxolation reaction compared to O$^-$. At the same time, hydroxides stimulate the inverse oxolation reaction resulting in species with lower nuclearity.[46] Highly alkaline HSILs thus form small oligomers with a higher average deprotonation per silicon, associated with (multiple) cations as the water content is low. We provide a simple model describing the charge per silicon in SI (Figure SI 6) and in the discussion below. As the Si concentration increases, so does the average nuclearity, while the monomeric ion gradually shifts to lower ppm, indicating increased shielding.

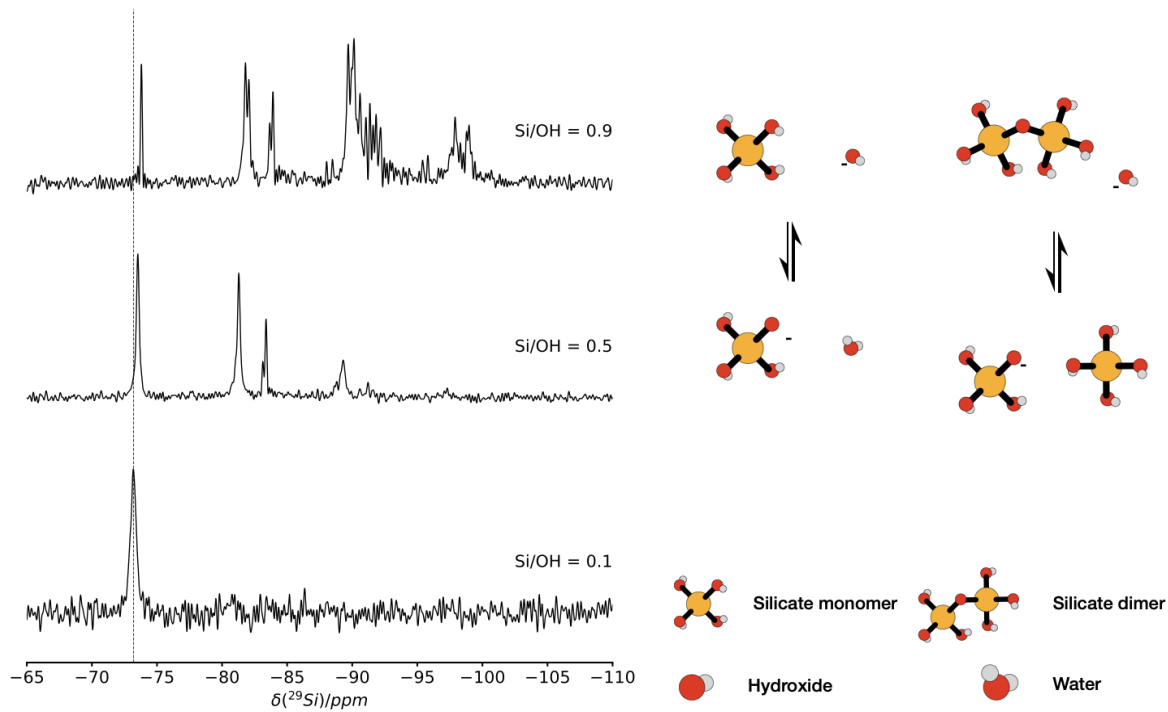

Figure 3 An increased alkalinity lowers the average nuclearity of silicates, while it increases the average charge per silicon center. The resulting increased shielding is evidenced by the gradual shift to higher ppm. The trends can be explained by the given reactions.[36] The figure is adapted from Asselman and coworkers.[37]

*Water*

The water content mainly affects the solubility of silicates. Dilution favors water molecules in the cations' hydration shells at the expense of silicates. Therefore, the cations are less involved in stabilizing the silicates in solution though ion association, so that the silicate solubility is reduced, and the formation of gel(-like) silicate particles is triggered.[18,19,46] For HSIL-based silicates containing traces of aluminate, earlier work reports a phase boundary at Si/OH = 0.5 and $H_2O$/MOH = 15, independent of the used cation.[19] Despite its effect on solubility, dilution does not significantly affect the speciation of the dissolved silicates for $H_2O$/MOH ratios below 20, as confirmed by $^{29}Si$ NMR investigations on HSILs featuring sodium[18] and potassium[38] as a charge-compensating cation, in a Si/OH ratio of 1 and containing traces of Al as well.

A recent computational study[17] highlights the occurrence of ion association using a model containing only monomeric single deprotonated silica with a Si/MOH ratio of 1. Although highly simplified, the study compares the liquid structure for different cations and confirms the same overall trend: in dilute, water-rich systems, ion solvation is dominated by water molecules, whereas silicates increasingly enter the solvation sphere when the water is getting deprived; a feature that is more pronounced for softer cations. Given the monomeric restriction of the model, this observation leads to (a combination of) two possible conclusions. (I) cations coordinate multiple silicate oligomers around themselves, creating favorable conditions for oligomerization. (II) oligomeric silanolates can coordinate with cations using not only their deprotonated silanols but also the protonated silanols and possibly their siloxane bonds, allowing a more intense interaction with the oligomers. Both interpretations can help to explain the structuring role of cations during the synthesis of silicates.

**MACROSCOPIC PROPERTIES**

Macroscopic parameters such as density, conductivity, and viscosity are often used as validation for molecular simulations. This section provides such experimental data to facilitate further simulation-based research. Once the link between these parameters is established, as is the case for ion association and conductivity for HSIL systems, the accessibility of macroscopic parameters allows a straightforward qualitative assessment of the microscopic parameters.

*Density and Viscosity*

The observed density profiles for HSIL-based silicate liquids having the composition $M_2SiO_2(OH)_2 \cdot y\ H_2O$, recorded at 25°C and having a water content (y) between 7 and 302, are presented in Figure 4a. Other temperatures are shown in Figure SI 2 and 3. For isoplethic liquids, the Cs-system displays the highest density, followed by K and Na, corresponding to the trend of their respective hydroxide solutions.[47] The values range between approximately 1 g/mL for highly diluted samples and approximately 2 g/mL for highly concentrated Cs-HSILs.

Dynamic viscosity measurements on the same compositions reveal the opposite trend with Na-HSIL being the most viscous. Viscosity is linked to the interaction energy of the liquid molecules, which in the case of highly ionic systems translates to ion association energy.[48–50] The higher viscosity of Na-HSIL compared to Cs-HSIL is thus related to the higher charge density of Na. A possible effect of the anionic silicates is excluded as the average connectivity remains roughly the same, as demonstrated above. Ionic liquids generally exhibit viscosities between 0.02 and 40 Pa.s, so HSILs reside on the lower end of the typical range.[48]

Viscosity was observed to be independent of the shear rate, indicating that HSILs can be considered Newtonian fluids (Figure SI 4). Interestingly, also Follens and coworkers found Newtonian behavior for highly diluted, zeolite-forming, silicate solutions.[51] Viscosity data at higher temperatures are given in Figure SI 3.

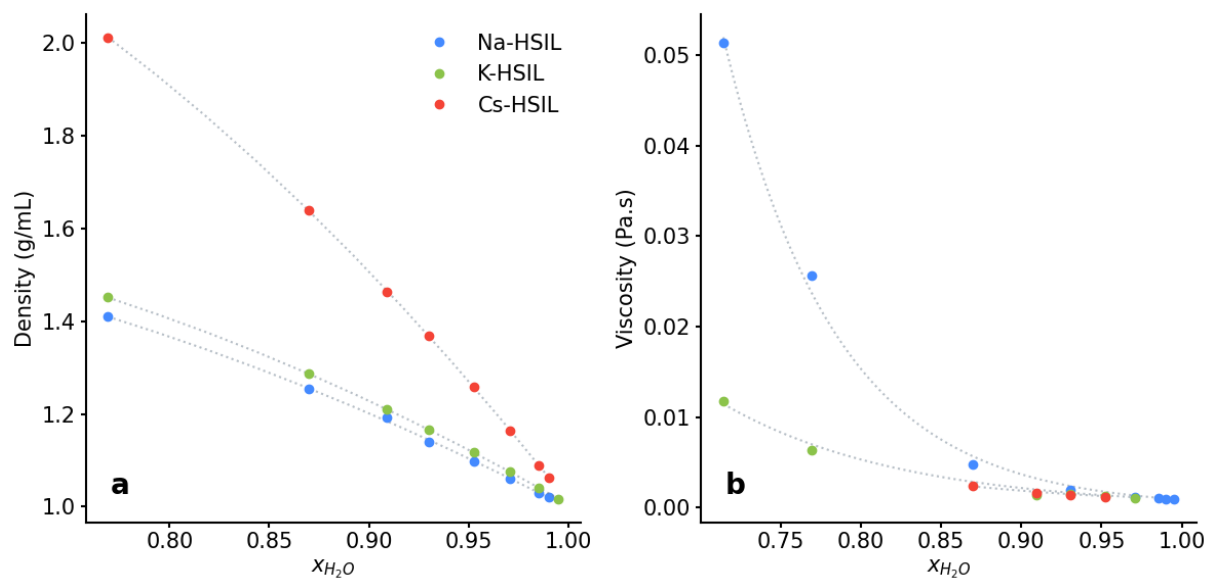

**Figure 4:** Density (a) and viscosity (b) data for HSILs having the composition $M_2SiO_2(OH)_2 \cdot y\ H_2O$ recorded at 25°C.

*Conductivity*

Ionic liquids typically exhibit a conductivity between 0.1 and 18 mS/cm at 25°C.[52] Distinctions are often made between protic and aprotic molecules. Protic ions can conduct charge not only through ionic diffusion but also via proton hopping (Grotthus effect), leading to increased conductivities.

Figure 5 displays the effects of dilution (a-c) and alkalinity (d-f) at 25°C. Conductivity values up to 50 mS/cm are achieved for K- and Cs-HSILs having a Si/OH of 0.125, which is somewhat higher than normally encountered for ionic liquids.[52] Dilution consistently impacts conductivity values, regardless of the cation or alkalinity involved, resulting in a characteristic profile for concentrated electrolytes.[18,53] In systems with low water content, ion association reduces the ion mobility and therefore the conductivity.[30] When this highly concentrated system is diluted, more ions become hydrated, reducing the number of associated ions and increasing conductivity. Further dilution shows a declining conductivity. In this part of the profile, the releases of associated ions no longer compensate for the decreasing charge density. For HSILs with a Si/OH ratio of 0.5, the conductivity peak occurs at around 10-15 water molecules, independent of the cation used. Also, alkalinity has a weak influence on this maximum, despite significant changes in speciation. Interestingly, the maximum closely corresponds to the occurrence of colloidal species, and to the maximization of the hydration shell of the alkali cations, typically coordinating about 15 water molecules contained in the first and second hydration shells.[19,54] Conductivity data for temperatures between 20°C and 90°C are provided in Figure SI 5. The conductivity increases with rising temperature.

Interestingly, increasing alkalinity always increases the conductivity when the water content is kept constant. This might be counterintuitive as a decreasing water content also increases the charge density but has a dramatically different effect. A simple model considering only orthosilicic acid and assuming the pKa values reported by Šefčík[55] shows the average charge per Si and the fraction of hydroxide ions that are not being consumed by deprotonation. Interestingly, compositions with Si/OH ratios above 0.5 show little to no available hydroxides. We can therefore assume that the anionic ion transport factor of the conductivity is fully governed by the silicates. After this critical ratio, the hydroxides are no longer fully consumed so they can contribute to the conductivity as well. As hydroxides are significantly smaller than silicate anions, they have a larger effect on conductivity. The model also indicates that the water content has only a minor impact on the charge/Si ratio. For instance, a sample with a Si/OH ratio of 0.1 shows a modest increase in its charge/Si ratio, rising from approximately 2.03 to 2.25, when the water/cation ratio decreases from 50 to 2.5 (Figure SI 6). We are positive that this model accurately describes the charge/Si and available hydroxides, even in systems with more complex speciation as the ionization constants are not strongly dependent on the speciation.[55,56]

The aforementioned molecular dynamics simulations capture the conductivity of samples with compositions of $MSiO_3(OH) \cdot y\ H_2O$ reasonably well.[17] While this allows a qualitative assessment of features like ion association, it also underscores the low impact of speciation on conductivity. It will thus be highly challenging to achieve a realistic speciation, solely based on MD-simulations. Additionally, the simulations only consider single deprotonated silicates, while the here-discussed model shows that charge/Si values above 1 are omnipresent, even for low alkaline, low water systems.

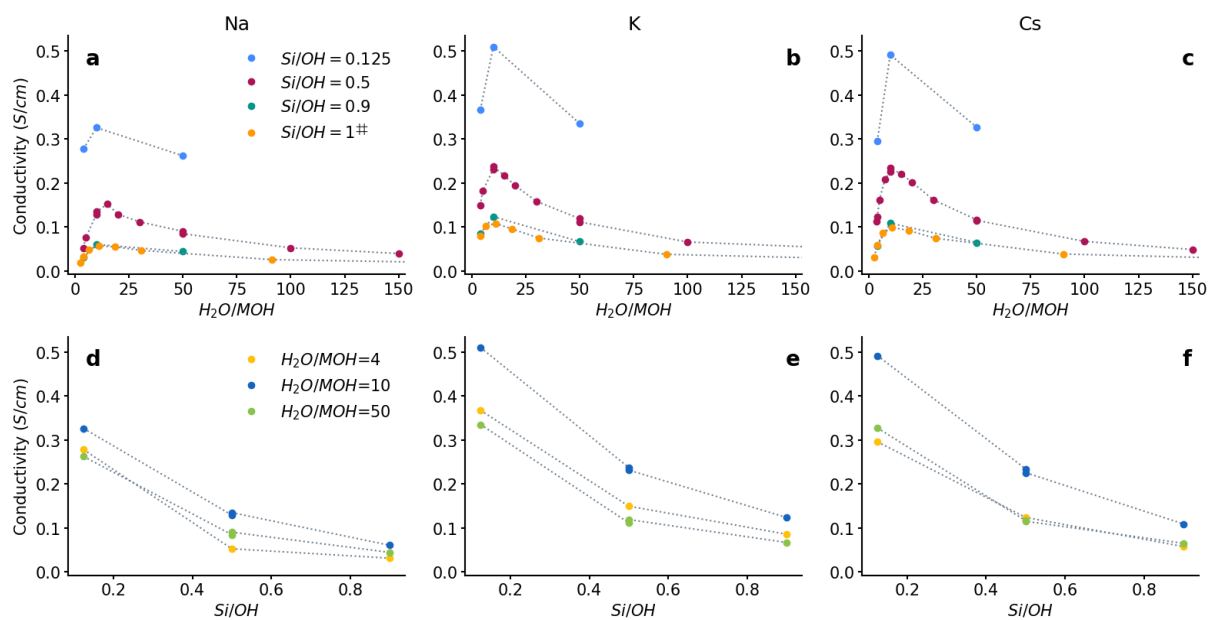

**Figure 5 shows the effect of hydration (a-c) and alkalinity (d-f) on the conductivity of Na-, K-, and Cs-HSILs. Data for Si/OH = 1 (♯), recorded by Vekemans and coworkers, is also included.[17]**

Finally, conductivity is often studied in relation to viscosity using the Angell-Walden criterion. This criterion compares the viscosity-molar conductivity relation with a 1M KCl solution, in an attempt to estimate how effects like proton hopping or ion association affect the molar conductivity. While the methodology has been proven to be arbitrary and in some cases inconsistent, it can still be used to qualitatively rank the conductivity of ILs for a given viscosity.[57] Plots are provided in SI.

**UNDERSTANDING CRYSTALLIZATION**

In a closed container, native HSILs resist crystallization. However, as the number of water molecules decreases, the system prompts the formation of glassy or crystalline alkali silicates. Crystallization can also be triggered by adding of heteroatoms such as aluminium or titanium under hydrothermal/ionothermal conditions, in HSILs that are often adjusted with additional alkali hydroxides and/or water. Up to this point, this research has focused purely on systems that typically form high-alumina zeolites as. these frameworks readily form without requiring hydrothermal conditions. There are preliminary indications that also all-silica materials could crystallize from HSILs, but their nucleation and growth typically requires a high temperature to overcome the activation barriers involved in the formation of Si-O-Si bonds.[58]

Compared to the conventional gel- and sol-based synthesis methods HSILs have the benefit of homogeneity. Heterogeneous systems complicate data analysis as they invoke 'false environments'. These are local gradients, which can hardly be probed yet can be crucial to understanding crystallization. In the case of zeolite formation, the occurrence of false environments is related to concentration gradients close to solid particles which result from the dissolution of these particles. HSILs offer a homogeneous alternative, reducing the chemical complexity, and thus facilitating data interpretation and model formulation. Additionally, as gel particles are maximally avoided, they are also not physically obstructing analysis.

This unique set of advantages has been explored in two recent communications where we explore the applicability of in situ NMR, conductivity monitoring, and X-ray scattering. Static $^{27}$Al NMR monitors the liquid-born Al content and speciation, while conductivity changes directly relate to changes in alkalinity invoked by aluminosilicate bond formation.[39,58] As these techniques only capture the liquid side, we explored synchrotron SAXS/WAXS.[26,37] With this technique, it is possible to differentiate between crystalline and glassy condensation, a feature we cannot capture by studying the liquid side alone.[59] Future work aims to fully integrate the liquid and solid perspectives. Also density and viscosity changes serve as indicators for zeolite formation.[51,60]

While this work focused on providing data for simulation studies, the same data can be used to qualitatively explain some zeolite-related observations. The conductivity and viscosity of highly alkaline sodium and potassium systems, for instance, hint at the degree of ion association. Crystallizing zeolites from these liquids at moderate temperatures yields SOD and EDI frameworks that include additional anions (hydroxides in this case).[61,62] A simplified deprotonation model (SI) confirms the presence of available hydroxides in these liquids, which could potentially be incorporated into the framework.

**CONCLUSION**

Hydrated Silicate Ionic Liquids (HSILs) expand the applications of ionic liquids for the synthesis of inorganic functional materials from serving exclusively as structure-directing agents to also serving as a source of framework-building units. They can be seen as room-temperature hypo-hydrated alkali silicate molten salts with controllable silicate speciation. This work summarizes the microscopic effect of alkalinity, hydration, and cation type and provides macroscopic data, focused on conductivity, which can be used to validate computational studies. It is clear from the provided data that alkalinity, hydration and cation type play a vital role to determine the HSILs' macroscopic properties. Additionally, strong correlations of these parameters, e.g. between cation type and water content, are likely. From a broader perspective, the development of HSILs fits into the endeavor to better understand zeolite crystallization, even though other (porous) silicates can be synthesized in HSILs too.

**ACKNOWLEDGEMENTS**

N.D, E.B and B.J acknowledges joined funding by the Flemish Science Foundation (FWO; G083318N and G0AC524N) and the Austrian Science Fund (FWF) (funder ID 10.13039/501100002428, grant ID 10.55776/I3680, project ZeoDirect, I3680-N34 and grant ID 10.55776/I6800, project MeMeZe, I6800). This work has received funding from the European Research Council (ERC) under grant agreement no. 834134 (WATUSO). E.B. acknowledges FWO for a "Krediet aan navorsers" 1.5.061.18N. NMRCoRe is supported by the Hercules Foundation (AKUL/13/21), by the Flemish Government as an international research infrastructure (I001321N), and by Department EWI via the Hermes Fund (AH.2016.134). The authors acknowledge the DUBBLE – The Dual-Belgian beamlines (ESRF, Grenoble) for experimental support at the BM26 beamline.